\documentclass[twocolumn,10pt]{article}
\usepackage{amssymb,color,graphicx}

\begin{document}

\twocolumn[
    \begin{center}
    {\large \textbf{IS PSEUDOSCIENCE THE SOLUTION TO SCIENCE LITERACY?}}\\

    \vspace{5mm}

     C.J. Efthimiou
    and
     R.A. Llewellyn\\
    Physics Department\\
    University of Central Florida\\
     Orlando, FL 32816
    \end{center}
          ]

\begin{abstract}
In EISTA03 the authors  described  an ambitious project they have
embarked upon at the University of Central Florida to help improve
public understanding of the basic principles of physical science,
topics often included in the general education program at many
colleges and universities. After the pilot course was very
successful, the authors were motivated to work harder on the
course and explore new directions. They thus decided to create
versions of the course, nicknamed flavors, whereby each flavor
used a particular genre or theme of movies: Action/Adventure,
SciFi, Modern Physics, Superheroes, Astronomy, Pseudoscience.

Among all flavors, the Pseudoscience flavor was especially praised
by the students. Contrary to passive attendance in the traditional
course, the class was full of passionate discussions driven by
scenes in popular movies such as \textsf{Sixth Sense}, \textsf{The
Others}, and \textsf{Dragonfly}. Almost every single person in the
class (of 90 students) would participate in the discussions and
defend his/her positions. These discussions often were passionate
and quite `heated'. Eventually, scientific analysis was presented
by the instructors who always maintained a skeptic attitude on any
extraordinary topic requiring extraordinary proof, thus explaining
how the scientific method works and what is acceptable and what
unacceptable in science. Amazingly, a fraction of the students
(not always the same students in all topics) did not always agree
with the rational explanations; often their decisions were
influenced by personal fears and biases or emotions. Their
thinking at times revealed the introduction of arbitrary
assumptions and acceptance of anecdotal statements.

Many reasons have been put forward by scientists why they do not
want to be involved in a fight against Pseudoscience. For example:
pseudoscientific beliefs are irrational and therefore it is
impossible to fight them with rational arguments; by even agreeing
to discuss Pseudoscience we accept it as an established
discipline; Pseudoscience is a background noise that does not harm
science. All these (and many other) reasons may contain pieces of
the truth, but they are not necessarily true. However,
Pseudoscience posses a continuous and, in fact, increasing threat
to our society. It is impossible to estimate exactly the
irreversible harm that will do in the future. It is the
scientists' duty to act now to eliminate or, at least, contain
this danger. Therefore, in a time when the science literacy of the
public has declined and the threat of pseudoscience has increased,
revising the traditional science course to present science through
the window of Pseudoscience might offer not only a way to restore
science literacy, but also a way to help society by eliminating
misconceptions and  attacking growing trends (astrology, remote
viewing, psychic readings, etc) that might harm (financially or
otherwise) innocent and trusting citizens.
\end{abstract}


\section{\textsf{Physics in Films}}

In \cite{EL} the authors have described  an ambitious project
embarked upon at the University of Central Florida to help improve
public understanding of the basic principles of physical science,
topics often included in the general education program at many
colleges and universities.  The project is built around the use of
a wide range of popular films to illustrate and serve as the basis
for discussions of physical science concepts.  One or two brief
clips are shown each day, augmented by live demonstrations and
other visual aids.  (The complete films are viewed by the students
at home as a part of their homework.)  While still a young project
(about 24 months old at this writing), it quickly became a success
\cite{success} and a topic of student conversation
\cite{UCFFuture}.

Student interest and performance in the \textsf{Physical Science}
course have both increased dramatically compared with the
traditional teaching mode, which we still use in some sections.

\section{The Idea of Flavors}

After the pilot course was very successfully tested in four
sections (with total enrollment of 800 students), the authors were
motivated to develop the course further and to explore new
directions.  The original pilot course included movies that were
selected to span the entire topical range of the standard Physical
Science course. In the selection of the movies no attention was
paid to the genre or the theme of the movie; eventually all the
movies used were action, adventure, and science fiction films.
Encouraged by the enthusiasm of the students, the authors decided
to take the project to the next level by considering possible
extensions of the course that would accommodate the curiosity of
every student and would satisfy the needs of every instructor.
Thus, we decided to create versions (packages)---nicknamed
\textit{flavors}---of the course whereby each flavor used a
particular genre or theme of movies. As a result, plans were
developed to create the following flavors, each using a different
film genre:
 \begin{itemize}
    \item {\bfseries Action/Adventure} that would use action and adventure
          movies;
    \item {\bfseries SciFi} -- science fiction movies;
    \item {\bfseries Superhero} -- superhero movies;
    \item {\bfseries Modern Physics}-- movies that enable the teaching of topics                                           from Modern Physics;
    \item {\bfseries Astronomy} -- movies that contain
          topics related to astronomy;
    \item {\bfseries Pseudoscience} -- movies that
          include pseudoscientific topics;
    \item {\bfseries Metaphysics} -- movies that touch on
           questions of metaphysical content.
 \end{itemize}
The reader will immediately recognize the instructors' intent to
break out of the traditional constraints and to teach the concepts
of \textsf{Physical Science} using media aids and illustrative
examples that not only go far beyond the usual approach, but also
have already been accepted by the students as a part of their
everyday experience.

\section{Pseudoscience Flavor}

An idea is named pseudoscientific if it contradicts accepted
scientific data, yet it is presented as scientific. We hasten to
avoid a possible misunderstanding. A mistake or error in the
presentation of a scientific fact does not signal pseudoscience;
in these cases the source of the mistake or error is limited or
due to incomplete knowledge and is not an intentional act. The
borderline of pseudoscience starts when misrepresentation of facts
or promotion of unverified claims are intentional.  When it comes
down to movies, it is very hard to accuse the industry of
promoting pseudoscience  in the guise of artistic freedom.
However, due to the great influence movies have on the
scientifically unsophisticated, we shall label (with our apologies
to the directors) movies that deal with topics that contradict
scientific facts as pseudoscientific.

\begin{figure}[h!]
\begin{center}
\includegraphics[width=1.5cm,height=2cm]{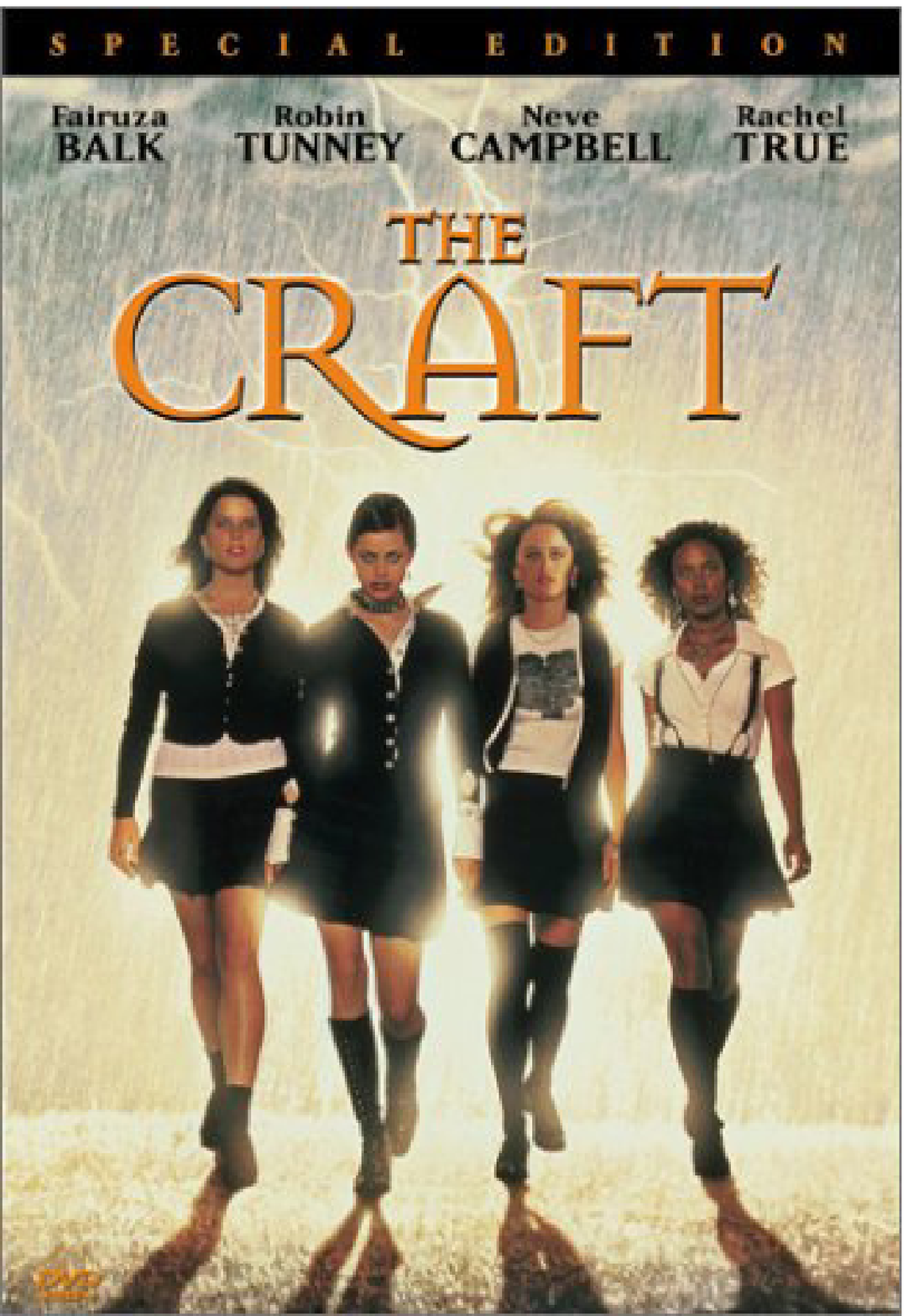}
\includegraphics[width=1.5cm,height=2cm]{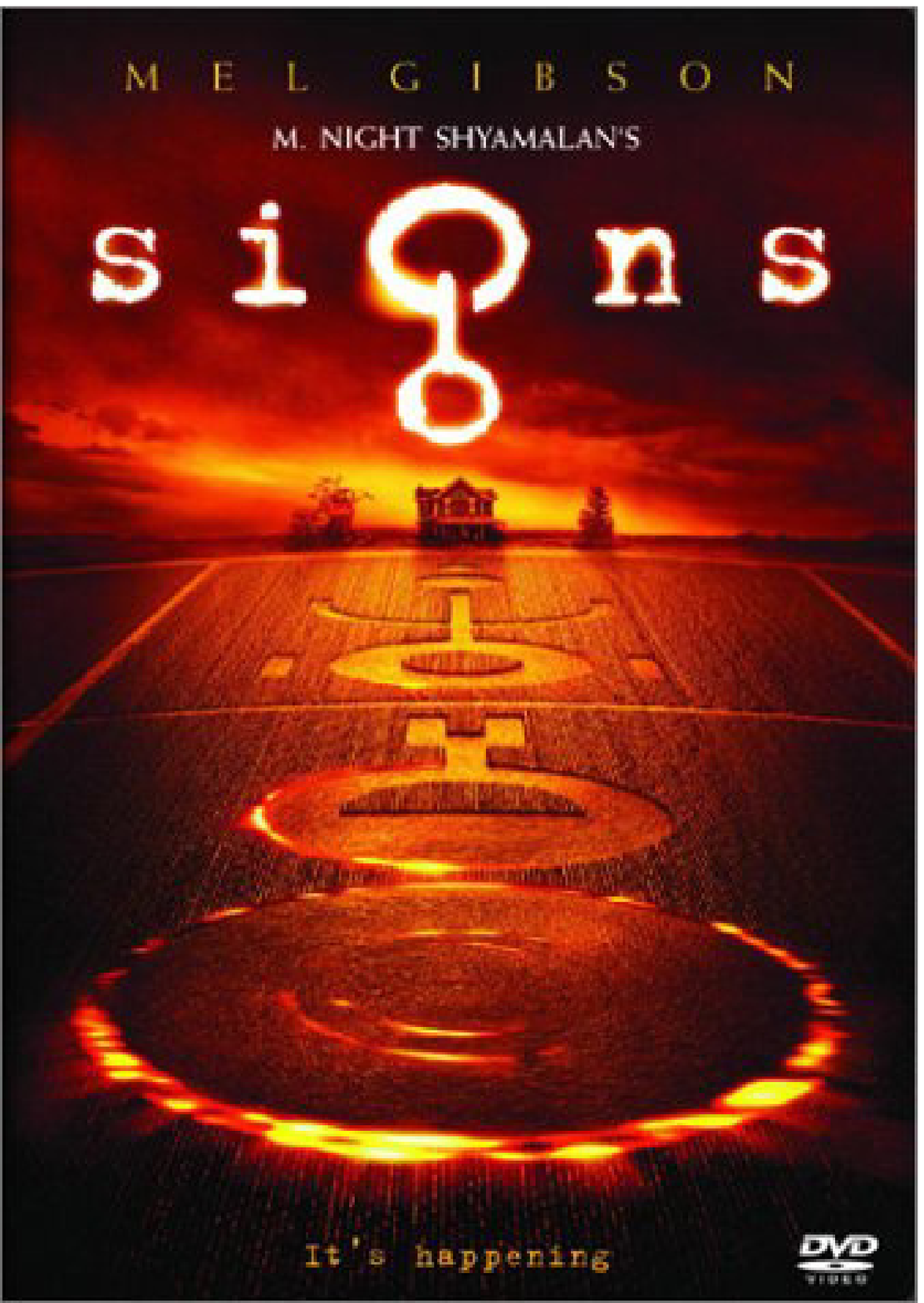}
\includegraphics[width=1.5cm,height=2cm]{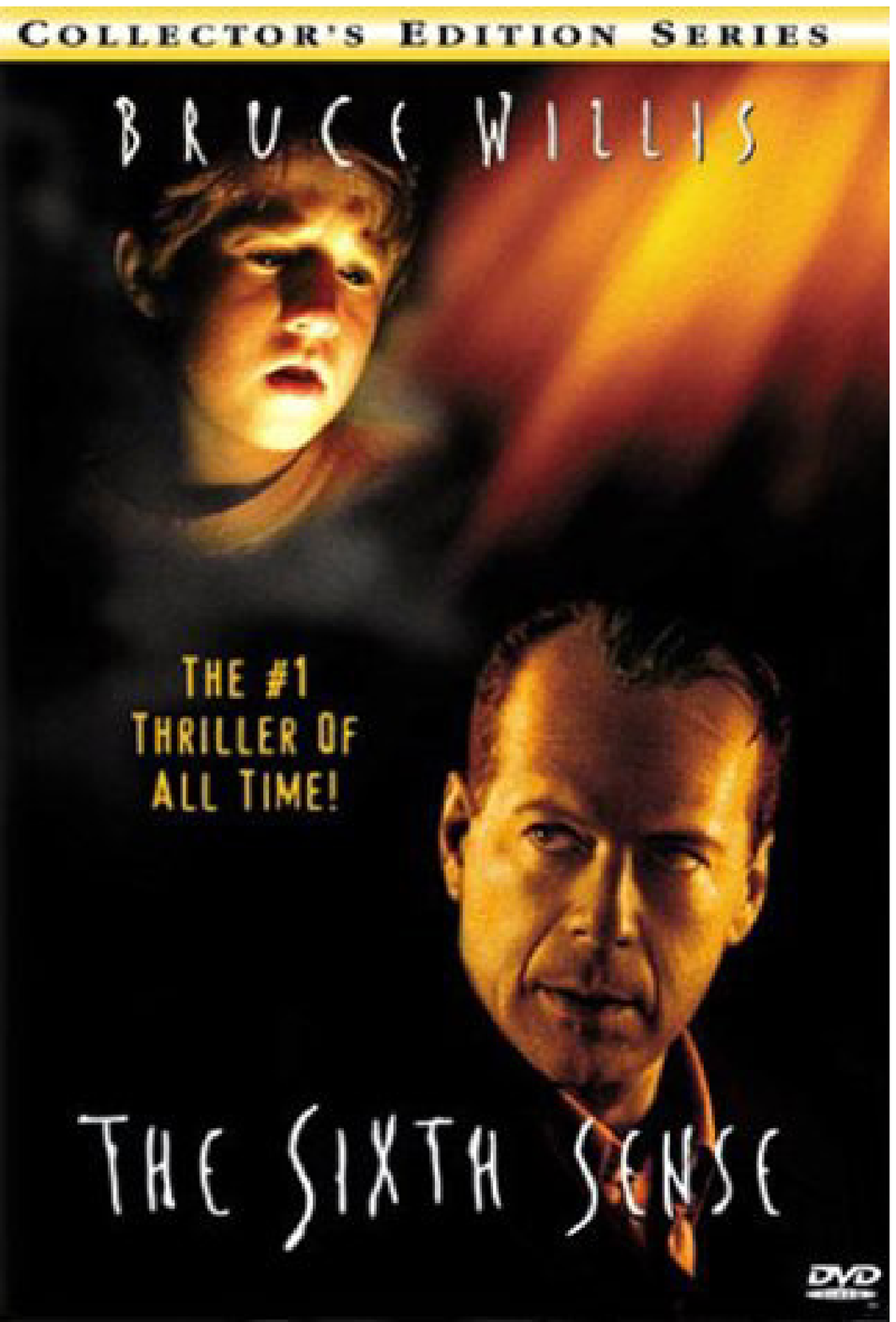}
\includegraphics[width=1.5cm,height=2cm]{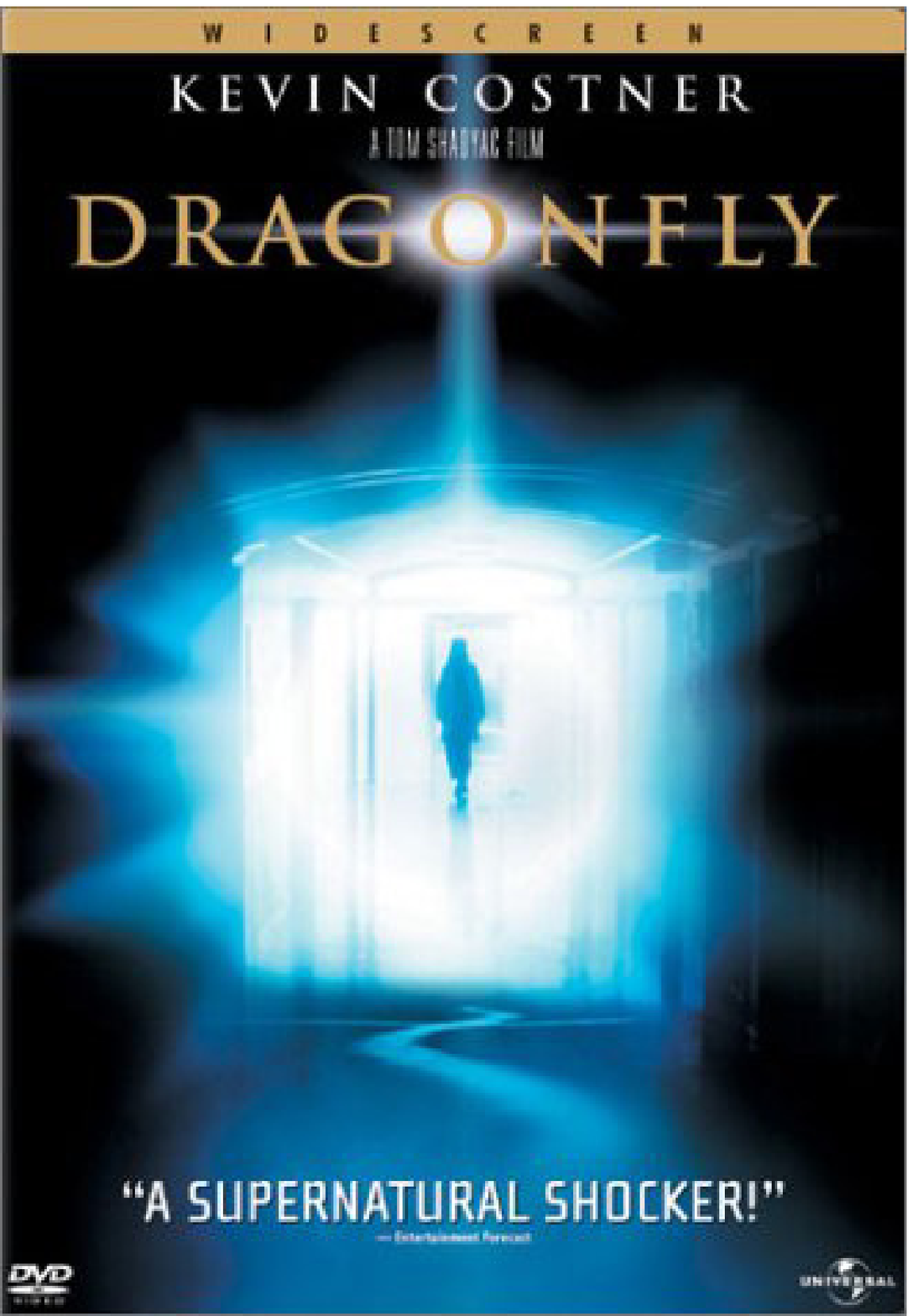}\\[1mm]
\includegraphics[width=1.5cm,height=2cm]{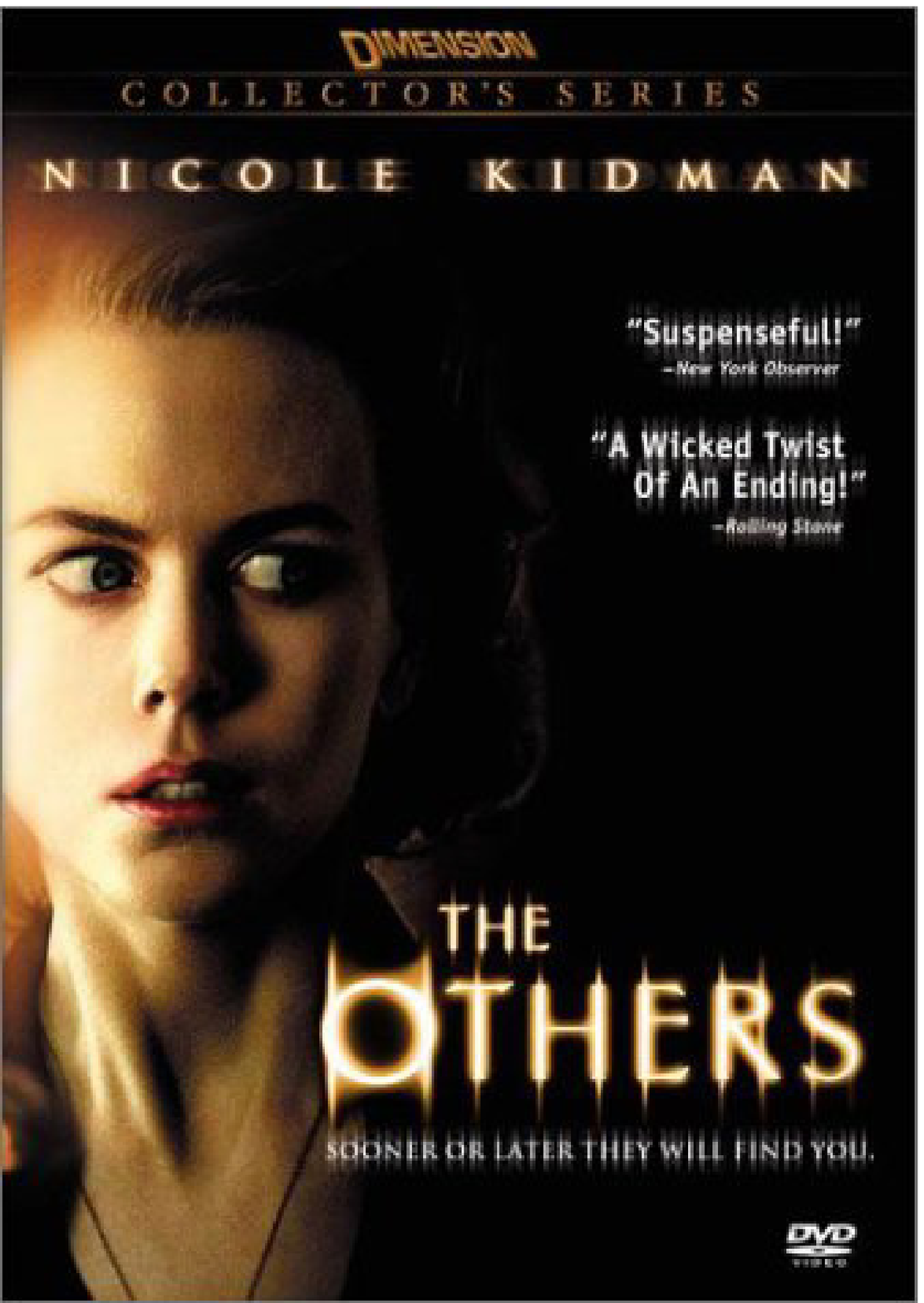}
\includegraphics[width=1.5cm,height=2cm]{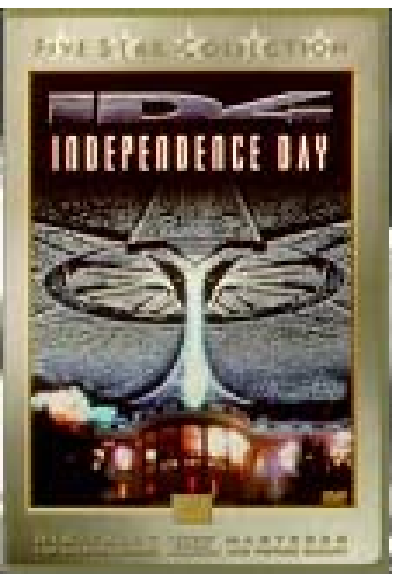}
\includegraphics[width=1.5cm,height=2cm]{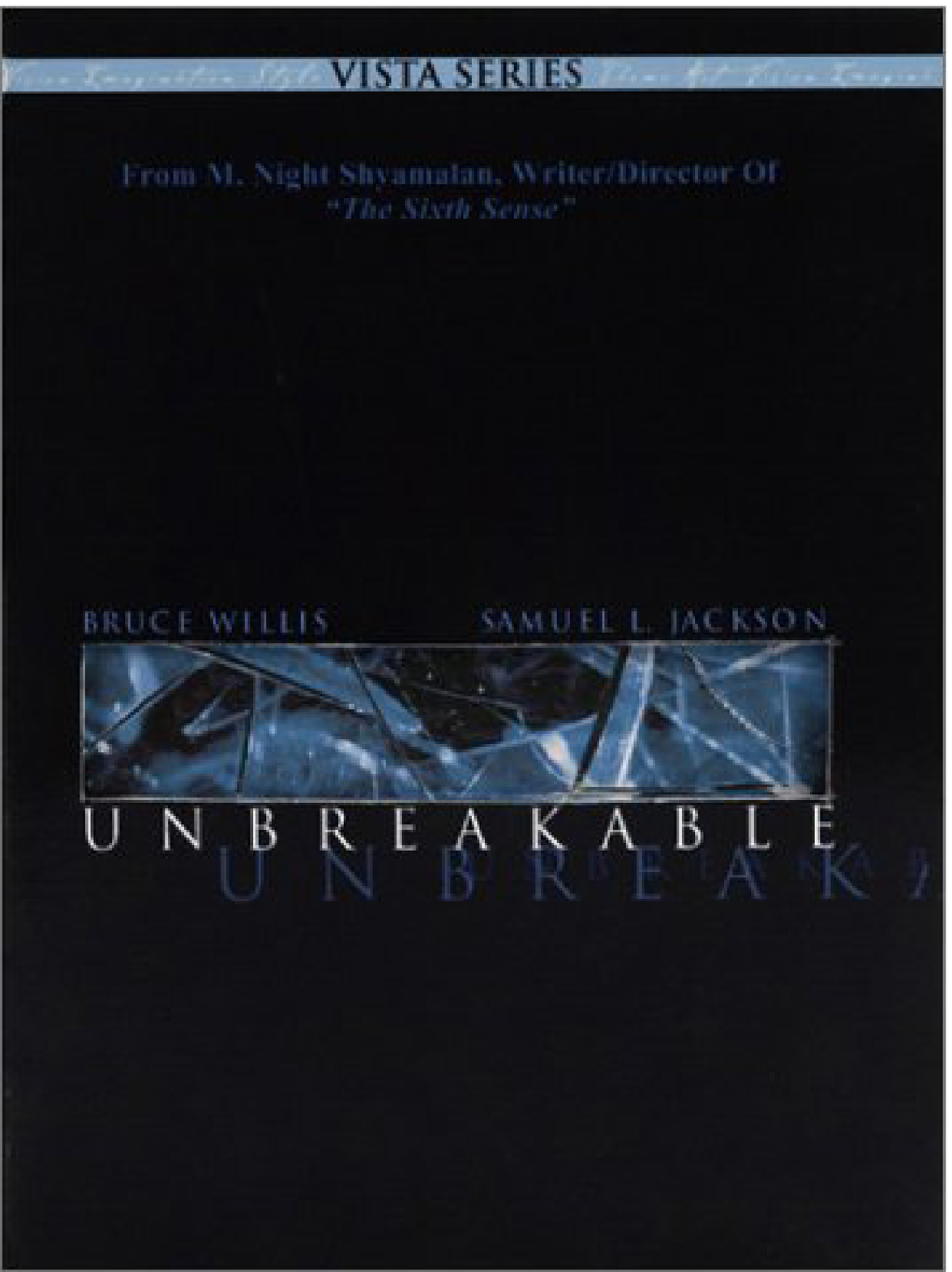}
\includegraphics[width=1.5cm,height=2cm]{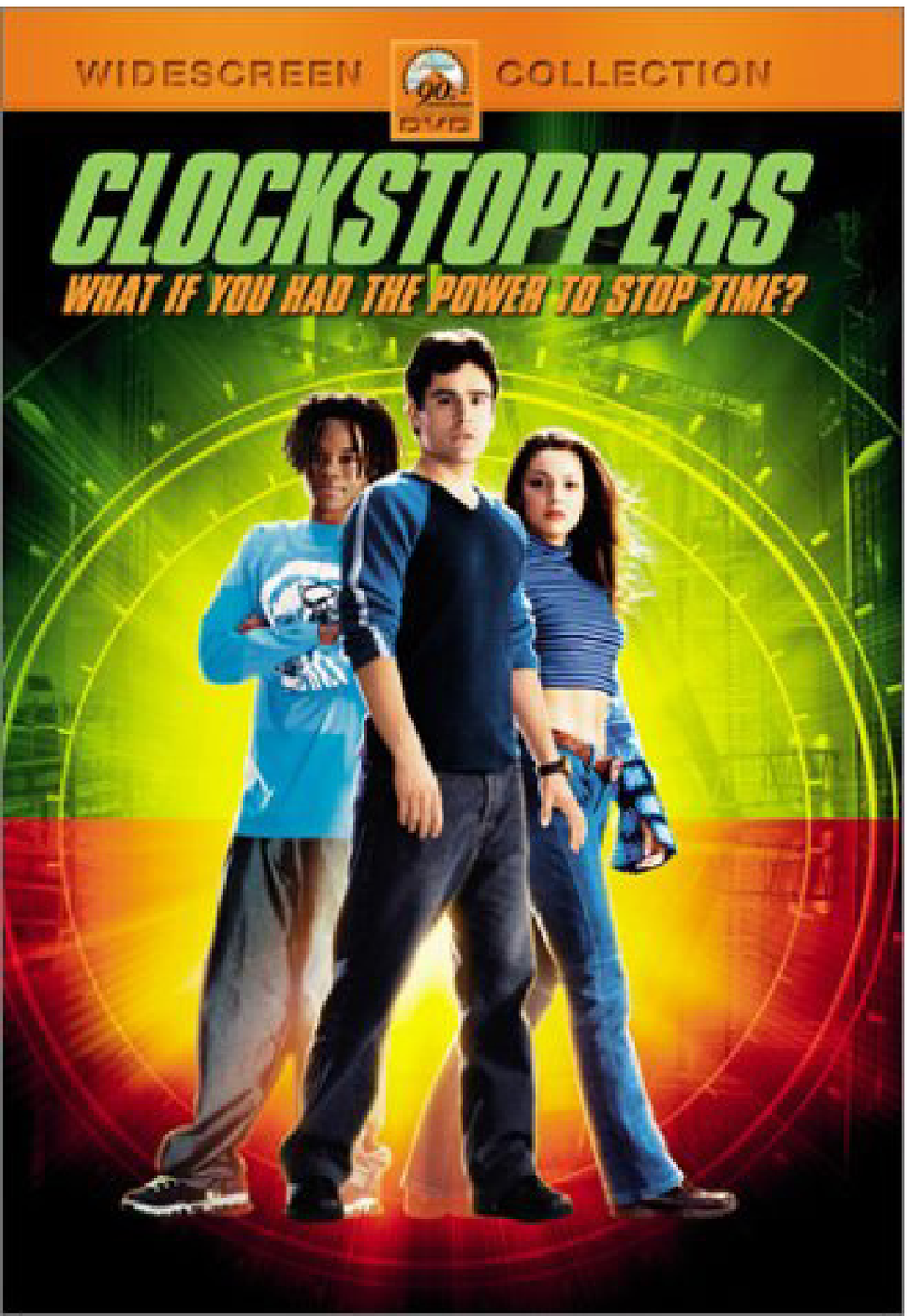}\\[1mm]
\includegraphics[width=1.5cm,height=2cm]{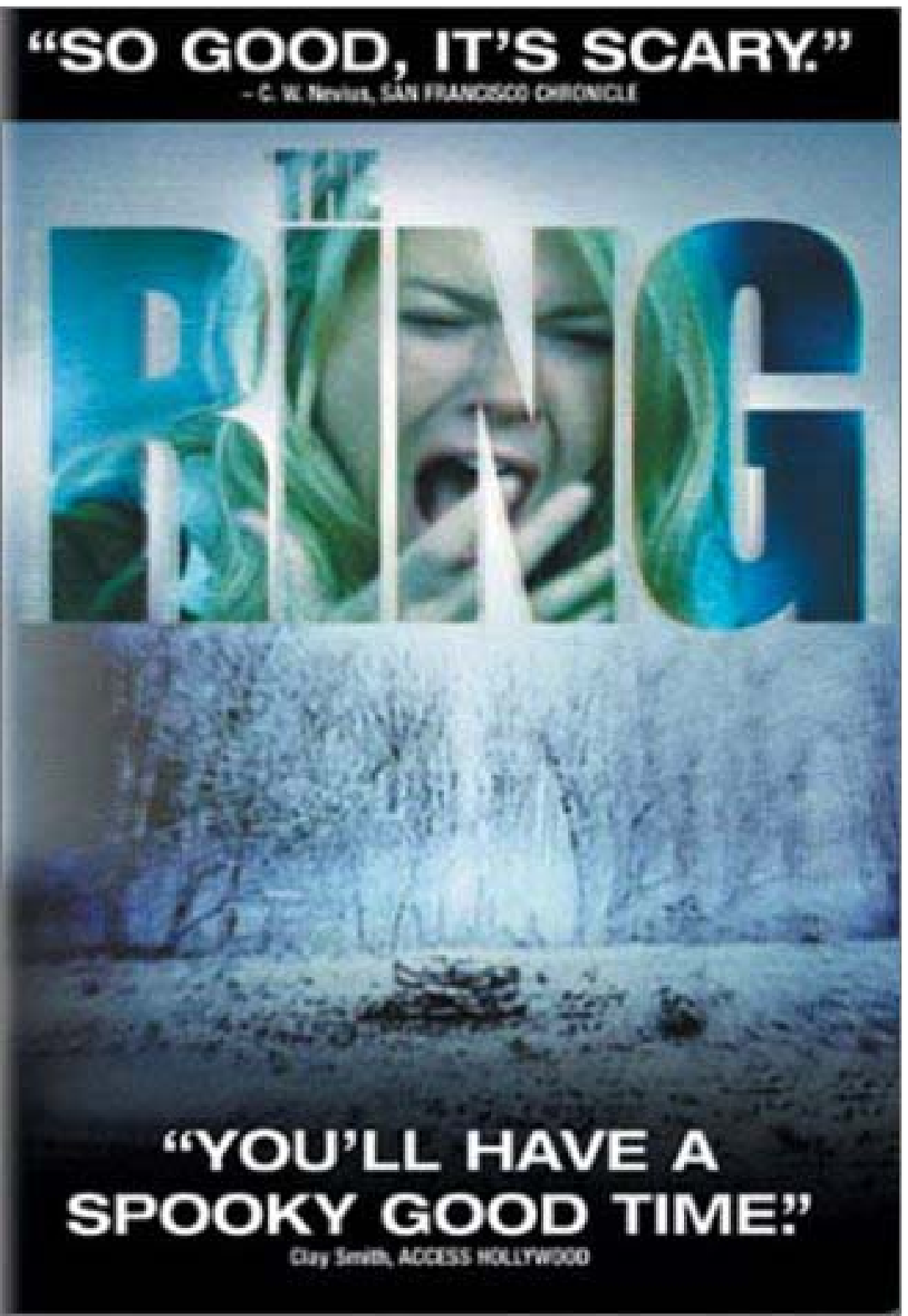}
\includegraphics[width=1.5cm,height=2cm]{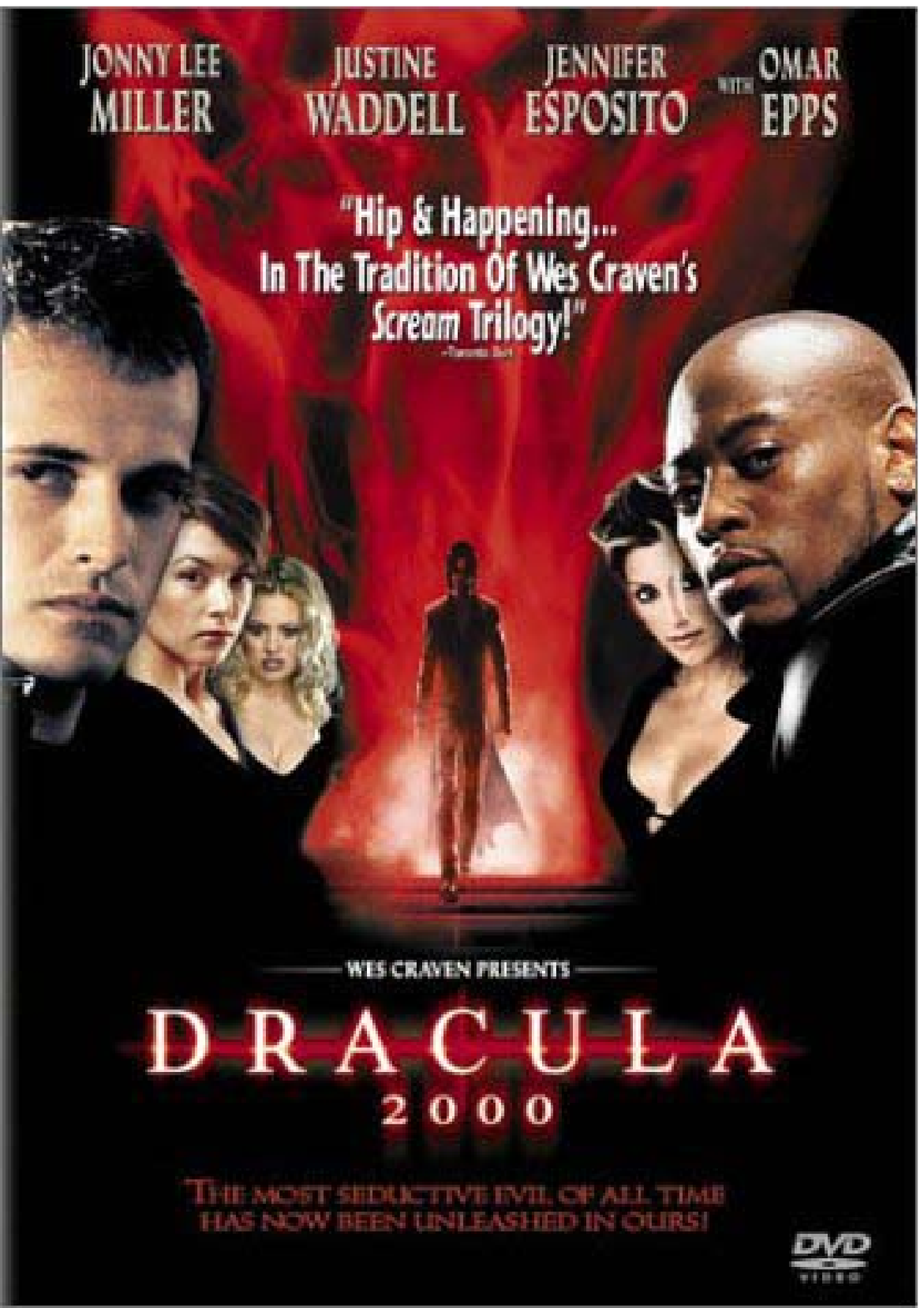}
\end{center}
\caption{Some of the films that are used in the
         \textit{Pseudoscience} flavor of the course.}
\end{figure}

There are numerous examples of pseudoscience that may be discussed
in  a class. However, within the schedule of a normal course only
a limited selection of topics is possible. In Summer 2003 we
taught a \textsf{Physical Science} class using the Pseudoscience
flavor and we chose topics that---we thought---the students would
find exciting and stimulating. The topics included:

\begin{itemize}
 \item intelligent life in the universe, alien visitors on Earth,
       alien abductions;
 \item fundamental interactions, ghosts;
 \item universality of the physical laws, magic;
 \item strength of materials, unbreakability;
 \item time, time reversal, time stopping;
 \item chemical reactions, zombies;
 \item physiology of human body, crossing to afterlife.
\end{itemize}

The instructors used clips from many Hollywood movies and
scientific documentaries. However, the required movies were:
\textsf{Clockstoppers}, \textsf{The Craft}, \textsf{Dragonfly},
\textsf{Independence Day}, \textsf{The Others}, \textsf{Signs},
\textsf{The Sixth Sense}, and \textsf{Unbreakable}.

The texts for this course were a selection of popular readings
\cite{popular} in pseudoscience: \textsf{The Demon-Haunted World},
\textsf{Did Adam and Eve Have Navels}, \textsf{Fads \& Fallacies},
\textsf{Why People Believe Weird Things} and one book that
describes how science works \textsf{The Scientific Endeavor}
\cite{Lee}.

\begin{figure}[h!]
\begin{center}
\includegraphics[width=1.5cm,height=2cm]{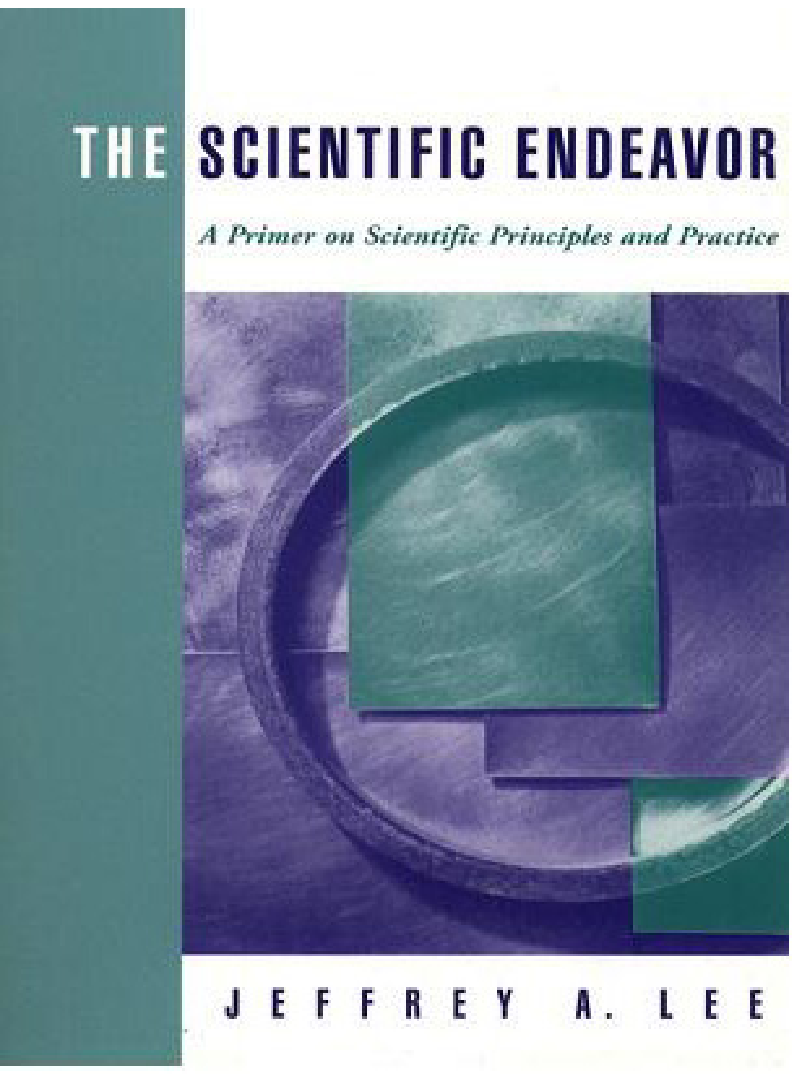}
\includegraphics[width=1.5cm,height=2cm]{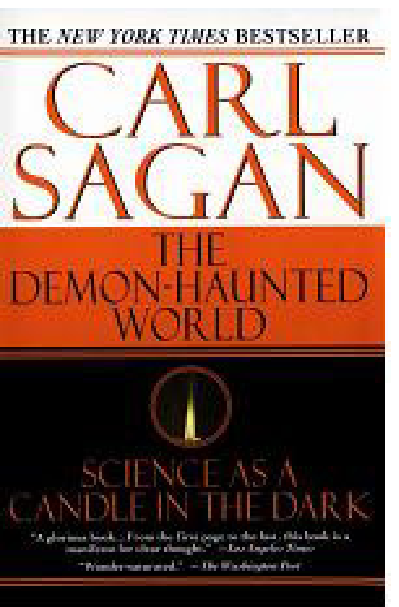}
\includegraphics[width=1.5cm,height=2cm]{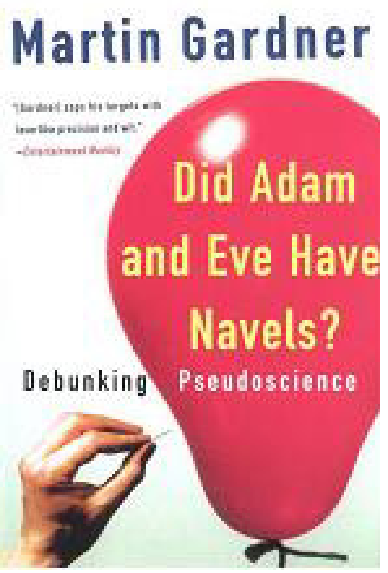}
\includegraphics[width=1.5cm,height=2cm]{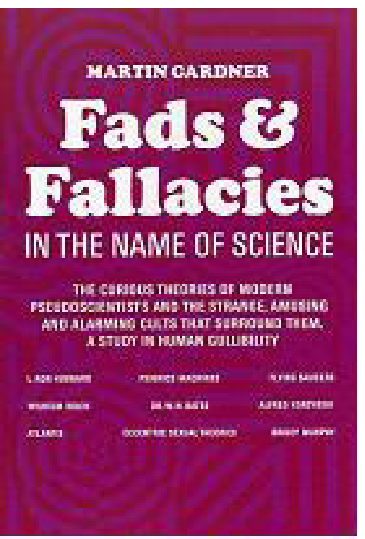}\\[1mm]
\includegraphics[width=1.5cm,height=2cm]{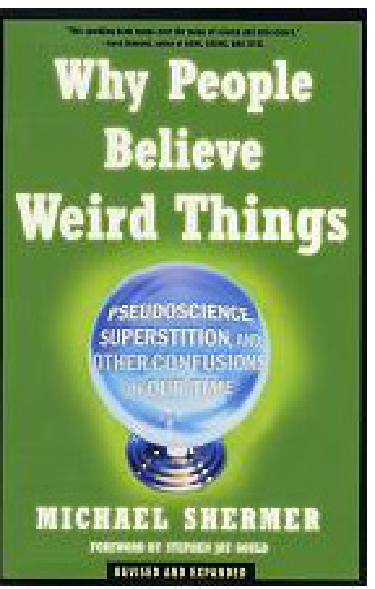}
\includegraphics[width=1.5cm,height=2cm]{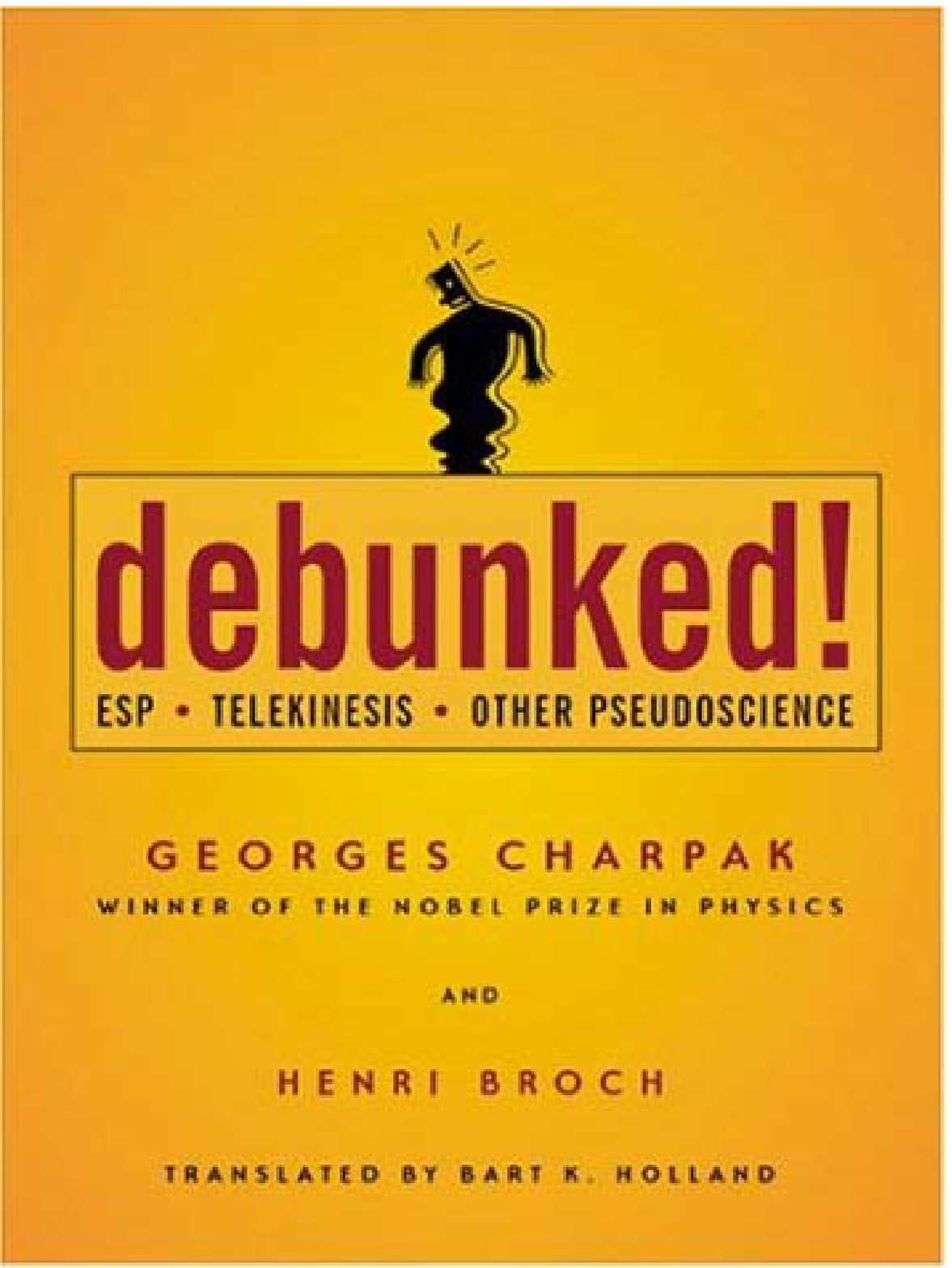}
\includegraphics[width=1.5cm,height=2cm]{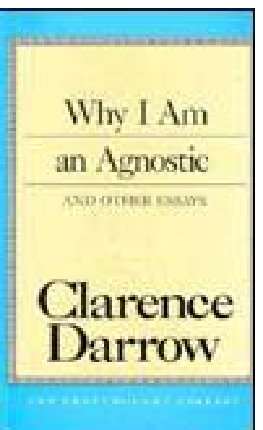}
\includegraphics[width=1.5cm,height=2cm]{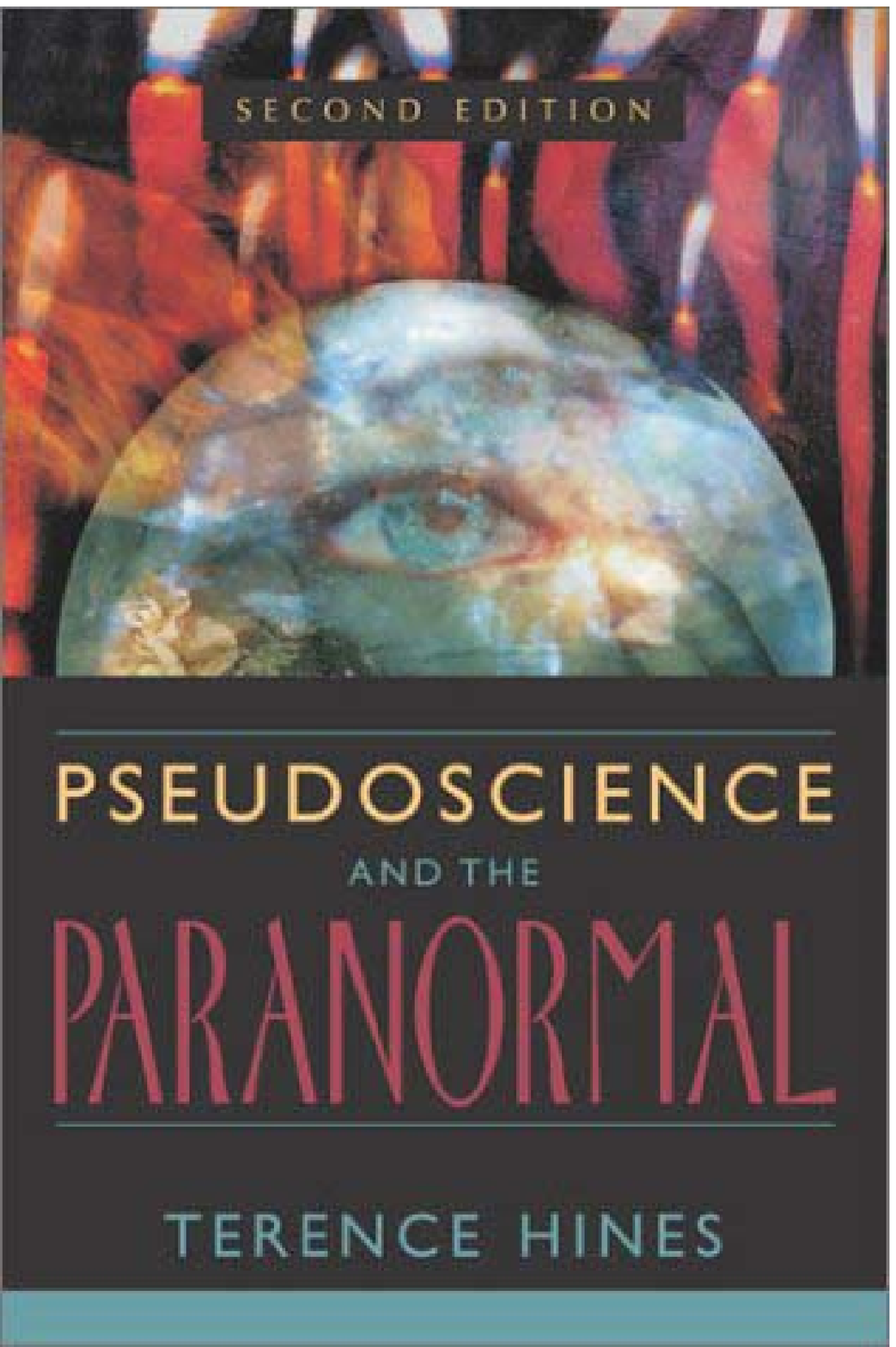}\\[1mm]
\includegraphics[width=1.5cm,height=2cm]{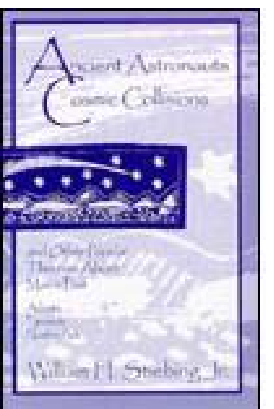}
\includegraphics[width=1.5cm,height=2cm]{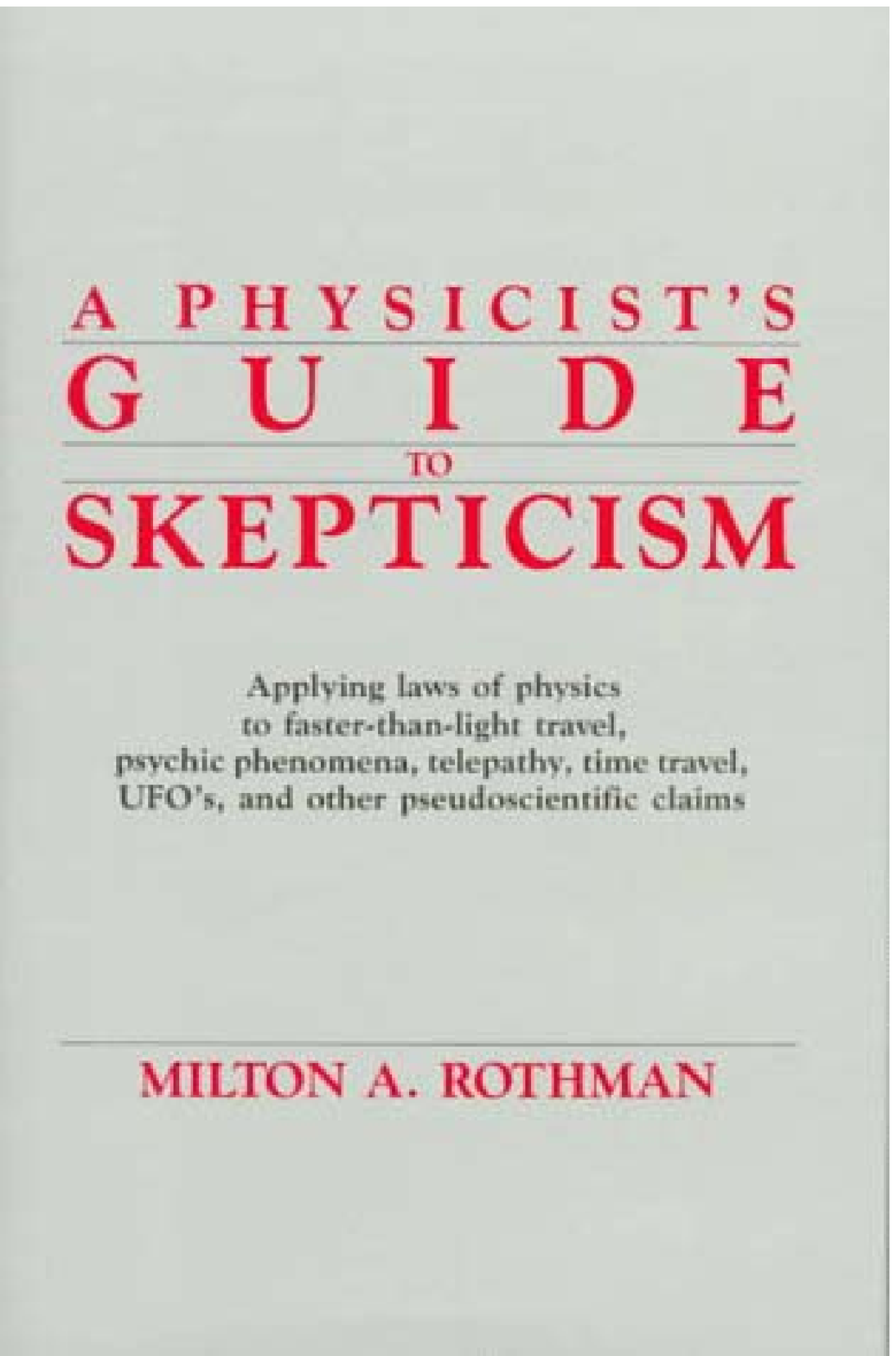}
\end{center}
\caption{Books which  have been  used or will be used
         as texts in the
         \textit{Pseudoscience} flavor of the course.}
\end{figure}

This summer we will teach the Pseudoscience flavor again (by the
time this conference is taking place, the class will be close to
its ending). \textsf{Dracula 2000} and \textsf{The Ring} will be
added to the required movies and several new texts \cite{texts2}
will be used and tested for their usefulness.

\vspace{5mm}
\hrule \hrule

\vspace{1mm}

\small
 \noindent {\bfseries EXAMPLE}: Among the topics discussed in physical science
is the concept of temperature.  Included are related topics of
heat transfer via conduction, convection, and radiation.  The film
\textsf{The Sixth Sense}  provides a background against which
temperature and heat transfer may be effectively discussed.

 \textsf{The Sixth Sense} is a film
 concerned with ghosts.  The movie consistently tells the viewer that ghosts
 like low temperatures, although why that should be is not explained.
In \textit{Chapter 26: Someone's in the kitchen}, the scene
clearly shows a sudden drop in the room temperature, so one
expects the appearance  of a ghost, and indeed one appears.

\begin{figure}[h!]
\begin{center}
\includegraphics[width=5cm]{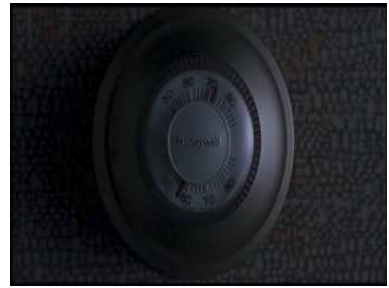}
\end{center}
\caption{In \textsf{The Sixth Sense}, the room temperature drops
before the appearance of a ghost. In this picture taken from the
motion picture, we see that the room thermostat has been set to
about $72^oF$ but the room temperature has dropped to $56^oF$.}
\end{figure}

To get a hint concerning the possibility that the appearance of
ghosts is heralded by a sudden drop in temperature, we shall look
at a case studied by scientists.

\begin{figure}[h!]
\begin{center}
\includegraphics[width=6cm]{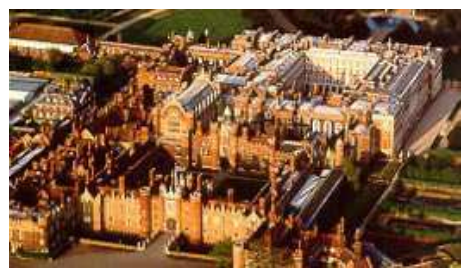}
\end{center}
\caption{Hampton Court Palace. The picture is borrowed from the
         official site \cite{HCP}.}
\end{figure}

In Hampton Court Palace \cite{HCP} near London, UK, there is a
well-known Haunted Gallery. It is said that the Gallery is stalked
by the spirit of Catherine Howard. Catherine Howard was the fifth
wife of King Henry VIII. King Henry executed her on February 13,
1542 for her indecent life. Many visitors to the room have
described strange phenomena in the gallery such as hearing screams
and seeing apparitions. Due to many reports of such occurrences, a
team of  psychologists, led by Richard Wiseman of Hertfordshire
University was called to investigate the claims
\cite{news1,news2}. The team installed in the Gallery thermal
cameras and air movement detectors. Then about 400 palace visitors
were asked if they could feel a ``presence" in the gallery. The
response was extraordinary: \textit{more than half reported sudden
drops in temperature and some said they sensed a ghostly presence.
Several people claimed to have seen Elizabethan figures}. The team
however discovered that the experiences could be simply explained
by the gallery's numerous old, concealed  doors. These exits are
far from draught-proof and the combination of air currents which
they admit cause sudden changes in the room's temperature. In two
particular spots, the temperature of the gallery plummeted down to
$36^o F$.
\begin{figure}[h!]
\begin{center}
\includegraphics[width=6cm]{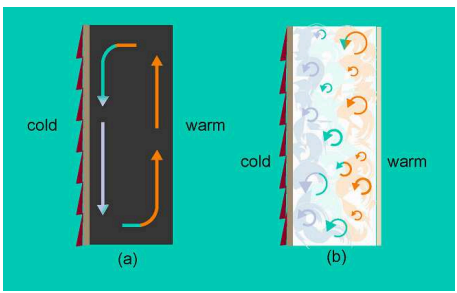}
\end{center}
\caption{Convection currents are created in any fluid which is
kept in a non-uniform temperature. The greater the non-uniformity,
the stronger the currents. The picture is borrowed from
\cite{Bolemon}.}
\end{figure}
``You do, literally, walk into a column of cold air sometimes,"
said Dr Wiseman. Convection is one of the three ways heat
propagates; the other two  are conduction and radiation.
Convection appears in fluids that have a non-uniform distribution
of temperature. As a result, currents inside the fluid will be
such as to attempt to restore a uniform temperature. These
currents are stronger when the non-uniformity is greater. In the
case of the gallery rooms, the convection currents would be felt
as cold drafts, similar to those experienced by someone who opens
the door of a refrigerator a hot day in summer.

Wiseman and his team have investigated other haunted places
besides Hampton Court Palace with similar results \cite{news2}.
Although, self-identified `psychics' claim that what people
experience is evidence of supernatural phenomena, physicists are
using Occham's razor to eliminate the possibility of anything
supernatural.

\normalsize

\vspace{1mm} \hrule \hrule

\vspace{5mm}

The previous example describes the interweaving of supernatural
claims and physics briefly, but demonstrates how such topics may
be related to physics concepts. In the example, the common belief
that ghosts live in cold places provides the opportunity to
discuss temperature and heat transfer (among other physics
topics). Besides educating students about physics, the course
offers a unique opportunity to attack widely held
misunderstandings and misconceptions. We will report on the topic
of pseudoscience more thoroughly in the future.

\section{Scientists vs Pseudoscience}

The majority of mainstream scientists avoid or ignore dealing with
pseudoscience. Some of the reasons used to explain this attitude
are tabulated by Friedman \cite{Friedman}:
\begin{itemize}
 \item For whatever reason, people want to believe in pseudoscience;
       therefore it is essentially a religion, and it is not the
       business of science to criticize religious beliefs.
 \item Pseudoscience is irrational; rational argument cannot counter it
       successfully.
 \item There might be something in the mass of
       pseudoscience that is correct but at present unknown to science.
       We would be foolish to attack pseudoscience, if even a tiny part of
       it proves out.
 \item Unlike creationism, astrology does not attack
       science itself. Live and let live.
 \item We have enough to do communicating science; there is no time to
       attack false science or correct inaccurate perceptions about
       the personalities of scientists.
 \item We would dignify pseudoscience  by mentioning it at all.
\end{itemize}
However, as scientists we have a moral obligation to society not
only to promote science for the well-being of humanity but also
for the well-being of the individual members of the society, too.
The following passage from Park (page 212 in \cite{Park})
summarizes this position well\footnote{Words printed in italics
may be replaced by those in the brackets.}:
\begin{center}
\begin{minipage}{2.8in}
 \emph{Voodoo science} [pseudoscience] is a sort of background noise,
annoying, but rarely rising to a level that seriously interferes
with genuine scientific discourse... The more serious threat is to
the public, which is not often in a position to judge which claims
are real and which are \emph{voodoo} [not real]. Those who are
fortunate enough to have chosen science as a career have an
obligation to \emph{inform  the public about voodoo science} [help
the public make the distinction].
\end{minipage}
\end{center}
Belief in pseudoscience indicates, besides lack of understanding
for the scientific method,  lack of critical thinking and this is
dangerous for the science policy in the nations, dangerous for the
democracy, and dangerous for the believer. Who can forget the
bodies of the 39 cult members in Rancho Santa Fe, California who
committed suicide in the belief that a UFO shielded behind the
Halle-Bopp comet would take them to heaven?

The question `who is responsible?' is hard to be answered. The
cause of this is quite complex and if we were to try to study it
carefully and thoroughly, we would be lost in a labyrinth of
intricate social, political, and economic factors. However, we
believe that the media, and in particular, the entertainment
industry, may be at least partially responsible for the large
numbers of people who believe in pseudoscience. There has been a
large number of very successful Hollywood films that have
glorified pseudoscientific topics  such as astrology, extrasensory
perception, alien invasions and alien abductions, ancient
interstellar travellers, ghosts, etc. Unfortunately not everyone
perceives such films as entertaining fiction and often many people
have extrapolated the ideas to the everyday life. This, combined
with the passive attitude of the scientists described above, has
magnified the problem and the public science illiteracy seems to
grow.

\textsf{Physics in Films} is a course that exactly attacks this
problem: the entertainment industry cannot present its ideas
unchallenged any more. Even more, the pseudoscientific topics are
paid special attention and a great effort is invested  to help the
public make the distinction between science and pseudoscience.

\section{Student Perception}

 How do the students perceive the Pseudoscience flavor of the
 \textsf{Physics in Films} course?
 The instructors  made
use  in class of an electronic  personal response system whereby
each student could respond immediately to questions posted by the
instructors, their responses being automatically recorded and
tabulated by an in-class computer. This system, besides its
pedagogical value for giving quizzes with immediate feedback and
scores, was used to obtain data on the student's feelings and
reactions to the course and to record attendance. The results for
few course-evaluation questions are shown in Tables
\ref{table:data1} through \ref{table:data3}. The notation in those
tables are as follows: SA=strongly agree, S=agree, NO=no opinion,
D=disagree, SD=strongly disagree.

\begin{table}[h!]
\begin{center}
\begin{tabular}{|c|c|c|c|c|}\hline
 SA & A & NO & D & SD \\ \hline
 42\% & 42\% & 16\% & 0\% & 0\% \\ \hline
\end{tabular}
\end{center}
\caption{Data on the question ``The topics selected from the
movies for physics analysis were interesting".}
\label{table:data1}
\end{table}

\begin{table}[h!]
\begin{center}
\begin{tabular}{|c|c|c|c|c|}\hline
 SA & A & NO & D & SD \\ \hline
 75\% & 20\% & 2\% & 0\% & 4\% \\ \hline
\end{tabular}
\end{center}
\caption{Data on the question ``The instructors should develop
         this course further since it is more interesting than the standard
         physical science course".}
\end{table}

\begin{table}[h!]
\begin{center}
\begin{tabular}{|c|c|c|c|c|}\hline
 SA & A & NO & D & SD \\ \hline
 68\% & 25\% & 5\% & 2\% & 0\% \\ \hline
\end{tabular}
\end{center}
\caption{Data on the question ``I would recommend to my friends
         that they take this course".}
\label{table:data3}
\end{table}

\section{Where do we go from here?}

Since we have been teaching \textsf{Physics in Films}, we have
made progress in improving student perceptions of physical
science: students no longer consider \textsf{Physical Science} to
be a waste of their time; they now feel that they learn something.
Their opinions are tabulated in Table \ref{table:data4}.

\begin{table}[h!]
\begin{center}
\begin{tabular}{|c|c|c|c|c|}\hline
 SA & A & NO & D & SD \\ \hline
 60\% & 32\% & 5\% & 1\% & 2\% \\ \hline
\end{tabular}
\end{center}

\caption{Data on the question ``I think I learned something from
         this course".}
\label{table:data4}
\end{table}

However, changing the public's overall perception of science is
not easy. More and longer term effort is necessary. Fear and
unreasonable dislike for science is deeply-rooted in the minds of
the students, as Table \ref{table:data5} illustrates.

\begin{table}[h!]
\begin{center}
\begin{tabular}{|c|c|c|c|c|}\hline
 SA & A & NO & D & SD \\ \hline
 25\% & 26\% & 8\% & 18\% & 23\% \\ \hline
\end{tabular}
\end{center}

\caption{Data on the question ``I do not like science and I do not
         want to read anything on science once I am done with this course".}
\label{table:data5}
\end{table}

On the strength of the positive results from the course
\textsf{Physics in Films} we believe we may have found a way to
address the problem of low scientific literacy among the public,
in general, and the abysmally low interest in physical science
among college non-science majors, in particular.  Among all
flavors, the \textsf{Pseudoscience} is distinctly viewed by the
students as by far the most exciting and it thus presents a unique
tool to facilitate science literacy. We plan to continue to
develop the course, explore its application to other science
disciplines, and find ways to disseminate it to interested
instructors throughout the nation.


\end{document}